# GSR Analysis for Stress: Development and Validation of an Open Source Tool for Noisy Naturalistic GSR Data

Seyed Amir Hossein Aqajari
University of California, Irvine
Department of Computer Science
saqajari@uci.edu

Emad Kasaeyan Naeini
University of California, Irvine
Department of Computer Science
ekasaeya@uci.edu

Milad Asgari Mehrabadi
University of California, Irvine
Department of Computer Science
masgarim@uci.edu

Sina Labbaf
University of California, Irvine
Department of Computer Science
slabbaf@uci.edu

Amir M. Rahmani
University of California, Irvine
Department of Computer Science
School of Nursing
a.rahmani@uci.edu

Nikil Dutt
University of California, Irvine
Department of Computer Science
dutt@uci.edu

## ABSTRACT
The stress detection problem is receiving great attention in related research communities. This is due to its essential part in behavioral studies for many serious health problems, including mental illnesses, such as depression, anxiety, and personality disorders—physical illnesses, such as high blood pressure, heart attacks, and stroke. There are different methods and algorithms for stress detection using different physiological signals. Previous studies have already shown that Galvanic Skin Response (GSR), also known as Electrodermal Activity (EDA), is one of the leading indicators for stress. However, the GSR signal itself is not trivial to analyze. Different features are extracted from GSR signals to detect stress in people—the number of peaks, max peak amplitude, to name but a few. In this paper, we are proposing an open-source tool for GSR analysis, which uses deep learning algorithms alongside statistical algorithms to extract GSR features for stress detection. Then we use different machine learning algorithms and Wearable Stress and Affect Detection (WESAD) dataset to evaluate our results. The results show that we are capable of detecting stress with the accuracy of 92 percent using 10-fold cross-validation and using the features extracted from our tool.

## 1. INTRODUCTION
The endocrinologist Hans Selye, a famous stress researcher, defined stress as the body's physiological reaction to any demands that have disrupted its mental or physical balance [1]. Stress factors fall into two different categories: (1) short-term/acute stress, and (2) long-term/chronic stress [2]. Acute stress occurs when an emotional pressure happens in the recent past or in the near future. An argument with a supervisor, a traffic jam, or watching a horror scene of a movie could be examples of acute stress. On the other hand, stress caused by long-term emotional pressures is chronic stress. An unsatisfying career could be an example of long-term stress. If the body is healthy and in good shape, it recovers very quickly from acute stress. However, chronic stress can impact various aspects of a person's life and it can cause a broad range of health diseases. Cardiovascular disease, cerebrovascular disease, diabetes, and immune deficiencies are examples of health-related diseases caused by long-term stress [3].

The two major systems that respond to stress are the hypothalamic-pituitary-adrenal (HPA) axis and the autonomic nervous system (ANS). They respond to stress as an effort to re-establish a steady state on a psychophysiological level [4], [5]. Changes in sweat gland activity is one of the activities involved in this process. Thus, Galvanic Skin Response (GSR), which is related to such activity, is considered to be a reliable indicator of stress. [5]–[7]

To understand how GSR works, it is helpful to take a look at the physiological characteristics of the skin described in [8]. Sweat glands are small tubular structures of the skin producing sweat. Our body has about three million sweat glands, which have different density across the body. We can find sweat glands in large numbers on the soles of the feet, the palms and fingers, and on the forehead and cheeks. They produce moisture through pores towards the surface of the skin, whenever they are triggered. When the balance of positive and negative ions in this secreted fluid changes, the electrical current flows more readily. This results in decreased skin resistance, or in other words, increased skin conductance. Galvanic Skin Response (GSR) is a term used for this change. GSR is also known as Electrodermal Activity (EDA), Skin Conductance (SC), Electrodermal Response (EDR), and Psychogalvanic Reflex (PGR).

Although one of the main purposes of sweating is thermoregulation, sweating is also triggered whenever we are exposed to any kinds of stimuli - emotionally loaded images, for example. This type of sweating is called emotional sweating.

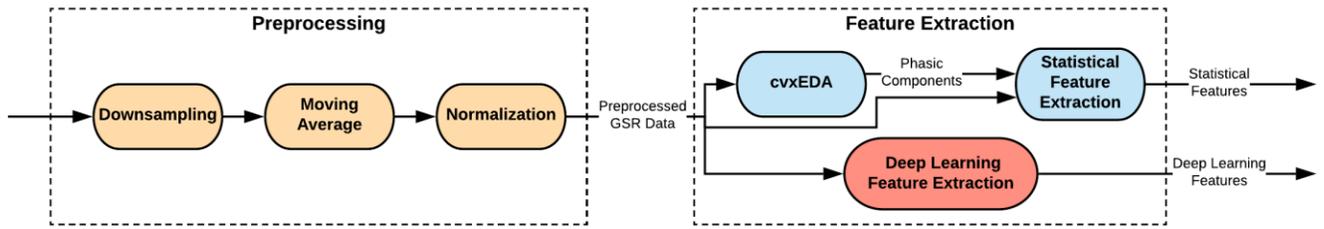

**Figure 1. Proposed Pipeline Architecture for GSR signal analysis**

Sweat secretion in our skin which reflects changes in arousal is driven unconsciously by our autonomic nervous system (ANS) in order to meet behavioral requests [8]. RespiBAN professional, Empatica E4, and Shimmer3 GSR+ are some examples of the devices that can be used to collect GSR signals.

According to [8], GSR signals consist of two main components: Skin Conductance Level (SCL) and Skin Conductance Response (SCR). The Skin Conductance Level(SCL) changes slightly on a time scale of tens of seconds to minutes. Depending on hydration, skin dryness, or autonomic regulation of an individual respondent, the rising and declining SCL is continuously changing. SCL, which is also called the tonic level of GSR signal, can differ significantly across different individuals. Due to this, the actual tonic level on its own is not that informative. SCR, which is also known as the phasic component of GSR signal, rides on top of the tonic changes and shows much faster alterations. Variations in the phasic component of a GSR signal are visible as GSR bursts or GSR peaks. The phasic component is sensitive to specific emotionally arousing stimulus events (event-related SCRs, ER-SCRs). These bursts can occur between 1-5 seconds after the onset of emotional stimuli. Quite the opposite, non-specific skin conductance responses(NS-SCRs) are not a consequence of any eliciting stimulus. These responses happen at a rate of 1-3 per minute spontaneously.

In this paper, we propose an open source tool in Python which gets noisy naturalistic GSR data and correctly detects GSR peaks that are a consequence of eliciting stimulus. Furthermore, one of the main novelties of our work is that we also use deep learning to extract new features from GSR data. Based on the experiments conducted in [9], the mean of GSR data, the number of peaks, and the max peak amplitude are the most discriminative statistical features for stress detection. Although their results show the importance of the mentioned features for stress detection, it does not prove that using only these statistical features are sufficient to create a stress model with the best accuracy. The biggest advantage of using deep learning for feature extraction is there is no need to choose what statistical features to extract. As a drawback, deep learning extracts a considerable number of features to achieve a good accuracy, which makes it less efficient in terms of runtime and memory especially for the large datasets. Therefore, providing a tool which can extract statistical features and deep learning features at the same time is highly valuable. In the end, we use the Wearable Stress and Affect Detection (WESAD) dataset [10] to evaluate our extracted features.

The rest of this paper is organized as follows. In Section 2 we discuss the related works. Section 3 describes our proposed pipeline architecture for analyzing GSR data. In Section 4 we discuss the preprocessing part of the pipeline. Feature extraction is described in Section 5. We evaluate our result using the WESAD data set in Section 6. Finally, Section 7 concludes the paper.

## 2. RELATED WORKS

The state-of-the-art in classification of GSR signals focuses only on the statistical features and lacks taking embedding extracted using deep learning algorithms into account. Soleymani et al. [11] developed a toolbox to extract features from different signals including GSR. The author built classifiers solely on statistical features including: amplitude and number of peaks, mean, and standard deviation of the signal.

Furthermore, there are toolboxes that provide integrated softwares. PhysioLab [12] and ANSLAB [13] are open source tools for EDA analysis, which are implemented in Matlab. These tools aggregate the information extracted from different signals including EDA. However, the feature extract module is limited to nonautomated statistical features.

Besides, researchers have implemented toolboxes for GSR analysis in Python [14]. These toolboxes also have the same limitation of the aforementioned studies. They just consider statistical features including: number of peaks, amplitude, rise time and decay time.

## 3. OUR PROPOSED PIPELINE ARCHITECTURE

Figure 1 shows our proposed GSR pipeline architecture for analyzing noisy naturalistic GSR data. There are two different stages in this pipeline: The preprocessing stage and the feature extraction stage. According to Figure 1, the preprocessing stage consists of three different modules. In the preprocessing stage, we process the data to make it suitable for feature extraction. Then, the feature extraction stage uses two different algorithms to extract the features from the preprocessed data. We use statistical algorithms and deep learning in our proposed architecture. In the following sections of the paper we explain each stage of the pipeline in detail.

## 4. PREPROCESSING

In this stage, we use down-sampling, moving averaging, and normalization to preprocess the data. At the end of this stage, a preprocessed GSR signal is ready and accessible for further analysis and features extraction. If the preprocessed data is only needed, it can be easily extracted from the output of the normalization module in the preprocessing stage from the source code.

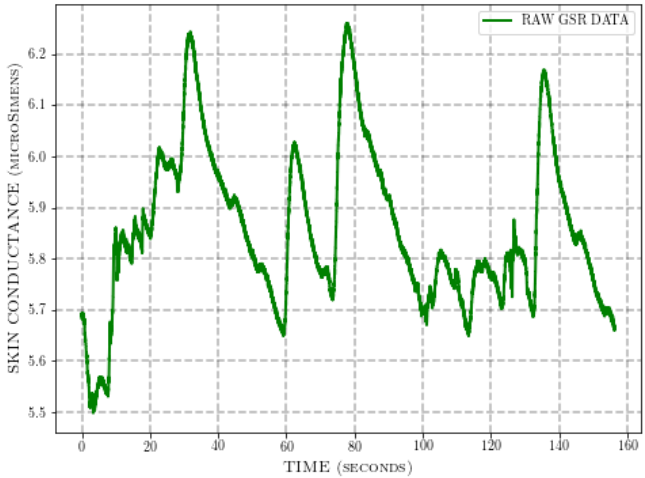 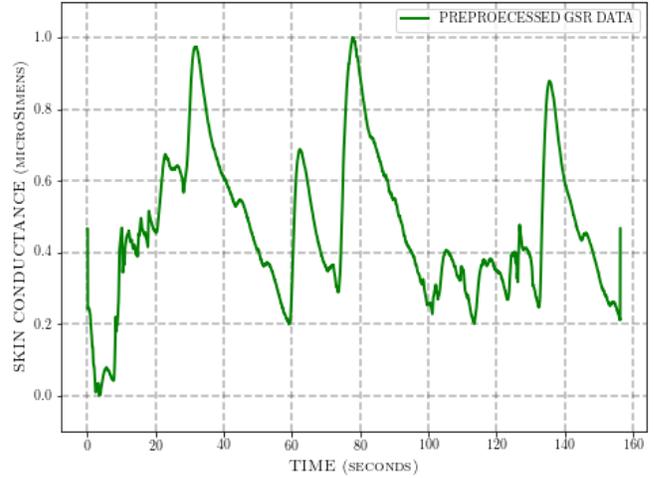

**Figure 2. RAW GSR data and GSR data after preprocessing**

## 4.1 Down-sampling

The GSR data is usually sampled at much higher frequency than needed. Therefore, down-sampling is done to conserve memory and processing time of the data without a significant risk of losing important aspects of the signal. In the preprocessing stage, the raw GSR data is down-sampled to the lower sampling rate. Based on studies conducted in [8], the GSR data can safely be down-sampled to 20 Hz or even less if the data originally was collected at 128 Hz.

## 4.2 Moving Average and Normalization

The raw GSR signal varies before or after a peak. This is due to individual differences in the tonic component of GSR or due to noise caused by movements or respiration artifacts [8]. After down-sampling the data, a moving average across a 1-second window was first used to smooth the data and reduce artifacts such as body gestures and movements, which are common in uncontrolled settings. Then, the smoothed data was normalized to reduce inter-individual variance using min-max normalization. Figure 2 shows an example of the raw GSR data collected with 128 Hz sampling rate from the Shimmer3 GSR+ sensor and the data after down-sampling to 20 Hz, moving averaging across a 1-second window, and min-max normalization.

## 5. FEATURE EXTRACTION

In this section we describe the second stage of the pipeline. One of the main novelties of our work is we extract deep learning features alongside the statistical features of the GSR data. At the end of this stage, deep learning features and statistical features are ready to be used for classification. In the next two following sub-sections, we discuss statistical and deep learning feature extraction modules of the pipeline.

## 5.1 Statistical Features

We select the number of peaks, the mean of GSR, and the max peak amplitude as three statistical features in our pipeline. According to [9], these three features are the most discriminative statistical features for stress detection. Calculating the mean of GSR is super straightforward. To calculate the other two features, we need to extract the GSR peaks that are caused by eliciting stimulus. A number of signal processing steps are required to derive GSR peaks, which are a consequence of eliciting stimulus [8]. (1) We need to extract the phasic data from the preprocessed GSR signal; (2) we apply a low-pass Butterworth filter on the phasic data to remove line noise. We used the typical cutoff frequency of 5 Hz divided by sampling rate of the data; (3) we need to identify onset and offsets; (4) to find the peaks, (a) We find the maximum amplitude value within each onset-offset window. (b) Subtract the amplitude value at onset from the max amplitude. (c) If the result (the peak amplitude) is higher than threshold, which is typically 0.005, we count it as a peak.

We can use the extracted peaks to calculate the number of peaks and max peak amplitude features.

### 5.1.1 Phasic Component Extraction

One of the major parts in analyzing the GSR signal is to correctly extract the phasic component of the signal from the original signal. In this paper, we used the cvxEDA algorithm to decompose the original signal into a sparse phasic component and a smooth tonic component. The cvxEDA algorithm is a novel algorithm presented in [15] for the analysis of EDA, also known as GSR based on maximum a posteriori probability, convex optimization, and sparsity. This algorithm has an excellent capability of properly describing the activity of the autonomic nervous system in response to affective stimulation, which was the main reason we picked this algorithm in our work. This model describes the recorded GSR as the sum of three terms: the phasic component, the tonic component, and an additive white Gaussian noise term incorporating model prediction errors, measurement errors, and artifacts. We used this algorithm to extract the phasic component of our preprocessed GSR signal for further analysis.

### 5.1.2 Onset and offset calculations

To calculate the number of peaks corresponding to event related SCRs, we need to find the maximum amplitude value within each onset-offset window in the original GSR signal [8]. In this section, we explain in detail how to extract onsets and offsets of the GSR signal from its phasic component.

Based on studies conducted in [8], specifications of onset and offset of the GSR signals are as follows. Onsets are all the points in which the phasic component of the signal crosses above the onset threshold, which is typically 0.01 μS. To find the corresponding offset of the computed onset, we check for the points in which the phasic component of the GSR signal crosses below the offset threshold, which is typically 0 μS. For each window, the time difference between its onset and offset should be above the duration threshold which is 1s. Any peaks before this duration threshold are considered as nonspecific skin conductance responses and should not be extracted.

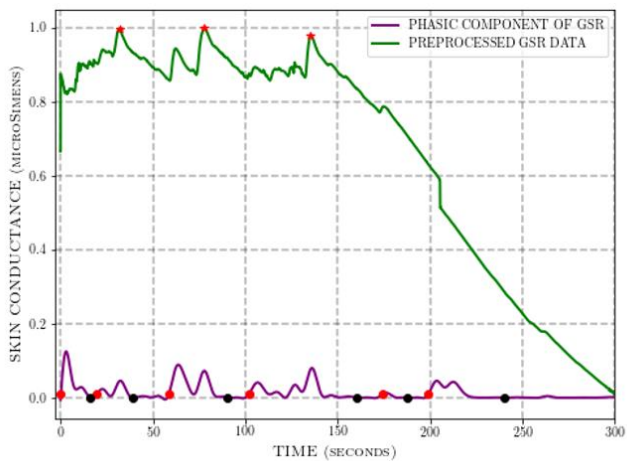 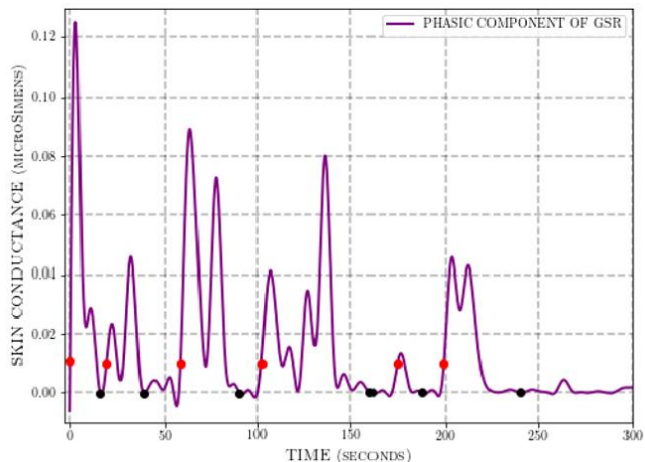

**Figure 3. GSR data and its extracted phasic component using cvxEDA algorithm**

Figure 3 shows the original GSR signal, along with the extracted phasic component. The extracted peaks are shown in the original signal with a red star symbol. Onset and offset of each window start with a red point as onset and finish with the black point as offset in the phasic component. Based on Figure 3, lots of peaks in GSR data are related to nonspecific skin conductance responses and are not extracted. This shows the importance of the extraction of ER-SCRs peaks based on the onset and offset of the signal, the phasic component of the signal, and threshold constraints.

## 5.2 Deep Learning Features

In [9], they show that the number of peaks, the mean of GSR, and the max peak amplitude are the most discriminative statistical features for stress detection. But, there might be some measurements of GSR data in which these three features are not the best selection of features for stress classification. Also, it is so hard and time consuming to pick the good selection of features among all the possible features in GSR data. With this aim, we also have the option of deep learning in our pipeline to extract the necessary features from the data.

In Deep Learning, Convolutional Neural Networks (CNN), as one of the discriminative models, was proposed by [16] for image recognition problems, where the model learns an internal representation of a two-dimensional input, in a process referred to as feature extraction. CNN is a hierarchical model consisting of convolutional, subsampling, and fully connected layers. The first two layers --convolution and subsampling-- carry out the feature engineering process. In this way, we overcome the tedious manual feature engineering process by humans.

To extract features from a CNN, we treat the pre-trained model as an arbitrary feature extractor, allowing the input signal to propagate forward, stopping at a pre-specified layer, and taking the outputs of that layer as our features. In other words, the output of the last subsampling layer, which is the input to the fully connected layer, is the features extracted within the CNN network.

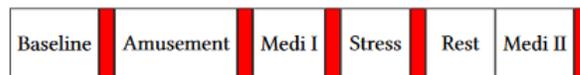
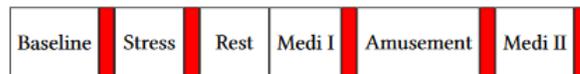

**Figure 4. Two study protocols used in WESAD dataset**

## 6. EVALUATION AND RESULTS
### 6.1 Machine Learning based Classification

We use machine learning based algorithms to evaluate the performance of our extracted features using the WESAD dataset. We use four different machine learning algorithms: (1) K-nearest-neighbor (kNN) with k between 1 to 10, (2) Naïve Bayes Gaussian classifier, (3) Random Forest with depth between 1 to 10, and (4) support vector machine (SVM). kNN method uses k number of nearest data-points and predicts the result based on a majority vote [17]. Naïve Bayes Gaussian classifier predicts the result based on the probabilities of each feature's Gaussian distribution [18]. SVM tries to find the best hyper-plan to divide the data points into different classes [19]. Random Forest classifier fits a number of decision tree classifiers on various sub-samples of the dataset and uses averaging to improve the predictive accuracy and control over-fitting [20]. We use the TensorFlow platform for classification and prediction. The TensorFlow is an open-source software library which is also used for machine learning applications such as neural networks [21].

### 6.2 WESAD dataset

WESAD is a publicly available multimodal dataset for wearable stress and affect detection [10]. In this data set, physiological and motion data are recorded from the Empatica E4 and RaspbiAN professional devices. There are 15 subjects during a lab study.

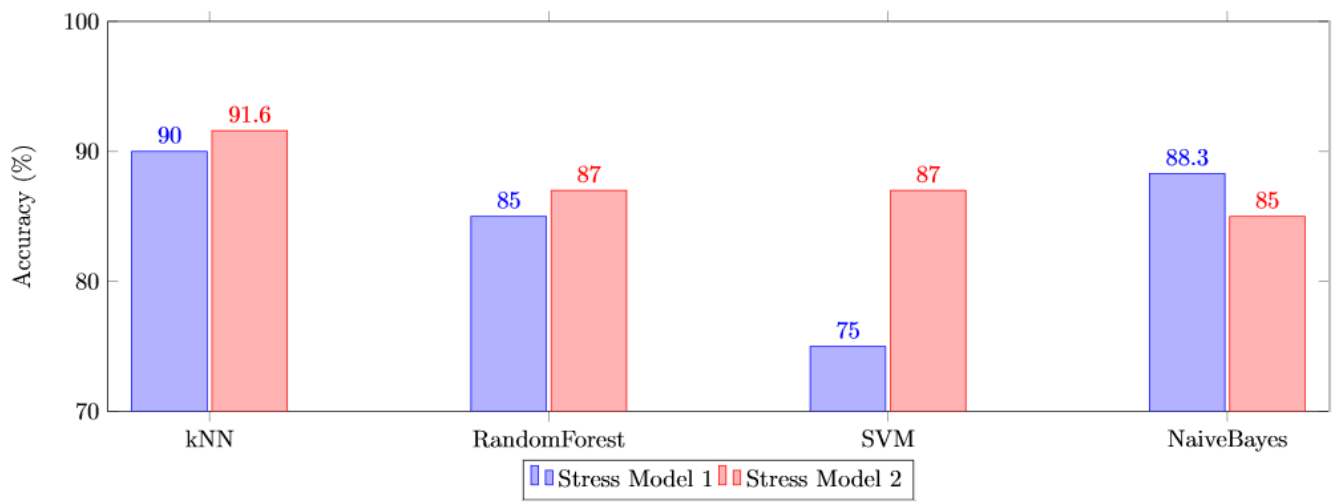

**Figure 5. 10-fold cross-validation accuracy on two different stress models using different machine learning algorithms**

The goal of this dataset is to elicit three affective states (neutral, stress, amusement) in the participants. There are two different versions of study protocol in this data set. Figure 4 summarizes these two protocols. They distinguish two different classification tasks based on these protocols. First, they define a three-class problem: baseline vs. stress vs. amusement. Second, they define a binary classification: baseline vs. stress. In this paper we are focusing on creating a binary classification to detect stress. We consider GSR data in the Baseline section labeled as not-stressed (0), and GSR data in the Stress section labeled as stressed (1) to create our model. Our model is created based on the GSR data collected from Empatica E4 wristband.

## 6.3 Results

We build our stress model based on four different machine learning algorithms (kNN, Naive Bayes, Random Forest, and SVM). In machine learning, irrevernt or partially relevant features can negatively impact model performance. Identifying the related features and removing the irrevernt or less important one is of great significance in order to achieve better accuracy for our model. With this aim, we train two different models based on the input features for each machine learning algorithm and report their accuracy. (1) The first model trains the data using only statistical features. (2) The second model trains the data using only deep learning features. We used 10-fold cross validation to evaluate the accuracy when the classifiers generalize across data points [22]. Cross validation also helps to prevent overfitting and achieve more information about our algorithms performance. Figure 5 shows the accuracy of the four different classifiers we used based on two different stress models. The best accuracy for the first model belongs to the k-nearest neighbor with k equals to 1, which is equal to %90. The best accuracy for the second model belongs to the k-nearest neighbor with k equals to 5, which is equal to %91.6. In [10], they could achieve the accuracy of %79.71 on the binary classification for EDA signals collected from Empatica E4 wristband. The results show that our proposed pipeline architecture is doing an excellent job in extracting related features for creating a stress model. (for both deep learning and statistical features.)

Based on our results, three out of the four machine learning algorithms achieve a higher accuracy while using deep learning features. Although this difference is not quite observable here, there might be some datasets in which this difference is noticeable. As we mentioned in the Introduction section, due to a lower number of statistical features compared to deep learning ones, classification using statistical features is much more efficient in terms of memory and runtime. Thus, having the both statistical and deep learning features in our tool, makes it extremely valuable.

## 7. CONCLUSION

We propose a new open-source tool in Python which not only does it correctly extract statistical features of GSR data that are a consequence of emotional stimulus, but also it extracts more new features using deep learning. To the best of our knowledge, there is no previous study claiming what statistical features of GSR data are the best features for stress classification. Our tool uses deep learning to automatically extract a considerable number of features of GSR data which are beneficial for stress detection. Although the large number of deep learning features for classification could cause an inefficiency in terms of memory and runtime, it could potentially achieve a higher accuracy compared to statistical features. Therefore, a tool which can extract both statistical and deep learning features of GSR data at the same time can facilitate and accelerate research in GSR signals analysis for stress detection.